\documentclass[aps,prd,reprint,twocolumn,preprintnumbers,floatfix,nofootinbib]{revtex4-1}
\usepackage[utf8]{inputenc}
\pdfoutput=1
\usepackage{graphicx}
\usepackage{amsmath}
\usepackage{amsfonts}
\usepackage{amssymb}
\usepackage{xcolor,soul}
\usepackage{epstopdf}
\usepackage{xcolor}
\usepackage{float}
\usepackage{subfigure}

\usepackage{relsize}
\usepackage{graphics}
\usepackage{hyperref}
\usepackage{mathrsfs}
\usepackage{amssymb}
\usepackage{booktabs}
\usepackage[normalem]{ulem}
\usepackage{amsthm}
\usepackage{cancel}
\usepackage{fontawesome}   
\usepackage{tabularx,ragged2e} % new

\hypersetup{
     colorlinks   = true,
     citecolor    = red,
     linkcolor    = blue,
     urlcolor     = blue,
}
\usepackage{physics}  %% DK
\newcommand{\bea}{\begin{aligned}}
\newcommand{\eea}{\end{aligned}}
\def\beq{\begin{equation}}
\def\eeq{\end{equation}}
\def\beqa{\begin{eqnarray}}
\def\eeqa{\end{eqnarray}}
\def\be{\begin{equation}}
\def\ee{\end{equation}}

\def\bse{\begin{subequations}}
\def\ese{\end{subequations}}

\def\tin{t_{\rm in}}
\def\Hin{H_{\rm in}}

\def\ain{a_{\rm in}}

\def\bea{\begin{eqnarray}}
\def\eea{\end{eqnarray}}

\def\Min{M_{\mathrm{in}}}

\graphicspath{{./figs/}}

\usepackage{pgfplots}
\pgfplotsset{compat=1.17}

\begin{document}
%\preprint{APS/123-QED}

%\thanks{}%

%\author{ }
%\email{ }
%\author{ }
%\email{ }
%\author{ }
%\email{ }

%\affiliation{$^a$ }
%\affiliation{$^b$ }
%\affiliation{$^c$ }

\preprint{}
\preprint{}

\vspace*{1mm}

\title{
%Improved Constraints on PBH and dark matter 
%accounting general relativistic accretion\\
Impact of general relativistic accretion on primordial black holes
}

\author{Santabrata Das$^{a}$}
\email{sbdas@iitg.ac.in}

\author{Md Riajul Haque$^{b}$}
\email{riaj.0009@gmail.com}

\author{Jitumani Kalita$^{a}$}
\email{k.jitumani@iitg.ac.in}

\author{Rajesh Karmakar$^{c}$}
\email{rajesh@shu.edu.cn}

\author{Debaprasad Maity$^{a}$}
\email{debu@iitg.ac.in}

\vspace{0.1cm}

\affiliation{
${}^a$Department of Physics, Indian Institute of Technology Guwahati, Guwahati, Assam, India
}
\affiliation{
${}^b$Physics and Applied Mathematics Unit, Indian Statistical Institute, 203 B.T. Road, Kolkata 700108, India
}
\affiliation{
${}^c$Department of Physics, Shanghai University, 99 Shangda Road, Shanghai, 200444, China
}

\begin{abstract} 
We demonstrate that general relativistic corrections to the accretion of relativistic matter onto primordial black holes (PBHs) can significantly enhance their mass growth during the early Universe. Contrary to previous Newtonian treatments, our analysis reveals that PBH masses can increase by an order of magnitude before evaporation, leading to substantial modifications of their lifetime and cosmological imprints. We quantify the resulting shifts in the minimum PBH mass constrained by Big Bang Nucleosynthesis (BBN), the revised lower bound for PBHs surviving today, and the dark matter parameter space allowed by PBH evaporation. Furthermore, we show that the enhanced accretion alters the high-frequency gravitational wave spectrum from PBH evaporation, potentially within the reach of future detectors. Our results provide a comprehensive, relativistically consistent framework to delineate the role of PBHs in early-universe cosmology and dark matter phenomenology.

%During their cosmological evolution primordial black holes (PBH) go through two important physical processes, namely accretion and Hawking evaporation. In the literature accretion in Newtonian approximation has been studied extensively and shown to have negligible impact on the PBH mass evolution. In this paper we revisit this analysis, and show that contrary to the earlier results the PBH mass can grow approximately ten times larger than their initial formation mass if general relativistic correction is taken into account in the accretion process. Such change in the mass leads to a series of improvement in the existing constraints on PBH parameter space. We obtain minimum PBH mass that can jeopardy the BBN to be $\simeq 6.9\times 10^{7}$ g, the expected PBH abundance in terms of particle like dark matter (DM) mass gets noticabe modification, the minimum possible mass that can survive till today turns out to be $\simeq 2\times 10^{14}$ g, the mass range within $10^{16}$ to $10^{21}$ g remain unconstrained to be the sole DM candidate. Finally we compute the modification in the high frequency gravitational wave spectrum emitted by the PBHs which can be detected in near future.      

\end{abstract}

\maketitle

%\tableofcontents

%{\color{teal} I have completed checking the manuscript except abstract and Introduction. For your reference, 
%\begin{itemize}
 %   \item Sentence(s) written {\orange orange} (is)are my suggestions, you are free to alter.
%    \item Strick out word(s) may be dropped.
 %   \item Underline word(s) create(s) confusion. Attention is required.
 %   \item Sentence(s) written {\red red} is(are) not clear to me. Strong attention is needed.
%\end{itemize}

%}

\section{Introduction}

Black holes (BHs) have been at the forefront of theoretical physics, bridging the realms of classical and quantum gravity. Traditionally, their phenomenological studies revolved around the gravitational capture of surrounding matter through the process of accretion in astrophysical environments \cite{1939PCPS...35..405H,1952MNRAS.112..195B}. However, a quantum mechanical perspective reveals that BHs can also emit particles through Hawking radiation \cite{Hawking:1975vcx}, an insight that reshaped the understanding of the very foundational nature of gravity and paved the way into its cosmological implications.

The groundbreaking discovery of gravitational waves by the LIGO and Virgo collaborations in 2016 \cite{LIGOScientific:2016aoc} further propelled BHs into the spotlight, intensifying their relevance in observational cosmology~\cite{2007coaw.book.....C}. In this realm, primordial black holes (PBHs) stand out as a potential key to unlocking several cosmological puzzles. These PBHs, formed from the gravitational collapse of overdensities in the early universe, span a wide mass range — from sub-gram scales to solar masses \cite{1974MNRAS.168..399C,1975ApJ...201....1C,Carr:2020gox,Auffinger:2022khh,Khlopov:2008qy}. Depending on their mass, PBHs can undergo both accretion and evaporation during cosmic evolution, influencing various phenomena such as dark matter (DM) production \cite{Masina:2020xhk,Hooper:2019gtx,Baker:2022rkn,Cheek:2021odj,Calabrese:2021src,Chaudhuri:2023aiv}, baryogenesis \cite{Baumann:2007yr,Fujita:2014hha,Datta:2020bht,Hook:2014mla,Barman:2022gjo,Hamada:2016jnq}, and cosmic microwave background (CMB) distortions \cite{Tashiro:2008sf,Acharya:2020jbv,Ricotti:2007au,Ali-Haimoud:2016mbv,Piga:2022ysp,Poulin:2017bwe,Serpico:2020ehh,Ziparo:2022fnc} to name a few.
Accretion is one of the pivotal processes of PBHs evolution in the early universe \cite{1972Ap&SS..15..153M,Harada:2004pf,Harada:2004pe,Lora-Clavijo:2013aya,Mack:2006gz,DeLuca:2020fpg,DeLuca:2021pls,DeLuca:2023bcr,Yang:2022puh,Yuan:2023bvh,Zhang:2023rnp}. While early studies on PBH accretion primarily focused on Bondi-Hoyle \cite{Ricotti:2007jk,Tabasi:2021cxo,Tabasi:2022fap, Stojkovic:2004hz,Jangra:2024sif,Nayak:2009wk,Park:2010yh,Wang:2023qxj,Manzoni:2023gon,khlopov:hal-03129467} models in the Newtonian framework, the general relativistic effect could be particularly important for PBHs that are embedded in a hot and dense environment of relativistic plasma in the early universe. 

%In fact all in the earlier studies the constraints on PBHs and their role in dark matter production were derived under considering the accretion in the Newtonian non-relativistic assumptions.

Motivated by this, we present a reformulation of the PBH accretion process within a general relativistic framework, and we systematically investigate its impact on the cosmological evolution of PBHs.
%---a regime that, to the best of our knowledge, remains largely unexplored, and investigate its impact on the evolution of PBHs for the first time.
% Motivated by this, we reformulate the PBH accretion process within a general relativistic framework, which {\orange remains largely unexplored to the best of our knowledge.}
% % , remains largely unexplored. to our knowledge, has not been explored in detail. 
% We investigate for the first time its impact on 3PBHs. 
Our analysis shows that relativistic corrections to the Bondi accretion model significantly enhance the accretion rate, leading to substantial mass growth of ultralight PBHs prior to their evaporation via Hawking radiation. By numerically solving the coupled equations governing accretion and evaporation in an expanding Universe, we derive updated constraints on the initial PBH mass spectrum, refine bounds on their abundance, and assess their viability as dark matter candidates. In addition, we predict a measurable shift in the high-frequency gravitational wave spectrum due to accretion-induced mass evolution. This unified treatment of PBH dynamics incorporating both evaporation and relativistic accretion provides a more accurate and testable framework for PBH based dark matter scenarios in light of current cosmological data.\\

\section{General relativistic accretion}
%Accretion is a process of the gravitational capture of the surrounding matter field by a compact objects. Simplest such process was first proposed by Bondi-Hoyle-Lyttleton (BHL) \cite{1939PCPS...35..405H,1952MNRAS.112..195B} with the assumption of spherical symmetry.

% We introduce here the basic formalism of general relativistic accretion in an axisymmetric asymptotically flat BH spacetime following \cite{Dihingia:2018tlr,Konoplya:2016jvv},

 We present the basic formalism of general relativistic accretion in an axisymmetric, asymptotically flat BH spacetime, following the approach outlined in \cite{Dihingia:2018tlr}, where spacetime is given by,
\bea
ds^2 =
g_{tt}dt^2+g_{rr}dr^2+g_{\theta\theta}d\theta^2+g_{\phi\phi}d\phi^2+2g_{t\phi}d\phi dt .
\eea
The horizon is defined as $g^{rr}=1/g_{rr}=0$ and the metric coefficients are functions of $(r,\theta)$. In this background, the dynamics of an ideal fluid is described by the conservation equations, $\nabla_{\nu} T^{\mu\nu}=0 ~;~ \nabla_{\mu} j^{\mu}=0$, where the indices $\mu$ and $\nu$ run from $0$ to $3$. The symbol $\nabla_{\mu}$ denotes the covariant derivative. Recall that any ideal fluid of energy density $\rho$, pressure $p$, and mass density $\tilde{\rho}$ is characterized by the energy momentum tensor $T^{\mu\nu}$, and the particle number current $j^{\mu}$, which are expressed as, 
\begin{align}
T^{\mu\nu} = (\rho+p)u^{\mu}u^{\nu}+pg^{\mu\nu}~~;~~
j^{\mu} = \tilde{\rho} u^{\mu} .
\end{align}
Here, $u^{\mu}$ is the four velocity in the lab frame supplemented by the condition $u_{\mu}u^{\mu}=-1$. 
% %$T^{\mu\nu}$ is the energy momentum tensor, $j^{\mu}$ is the particle number current and
% The symbol, $\nabla_{\mu}$ is the covariant derivative.
% %Expression for $T^{\mu\nu}$ and $j^{\mu}$ are
% %\begin{align}
% %T^{\mu\nu} &= (\rho+p)u^{\mu}u^{\nu}+pg^{\mu\nu} \\
% %j^{\mu} &= \tilde{\rho} u^{\mu} 
% %\end{align}
Defining a projection operator as $h^i_{\mu}=\delta^i_{\mu}+u^i u_{\mu}$, that satisfies $h^i_{\mu}u^{\mu}=0$, we project the energy momentum conservation equation into a three vector equation as,
\begin{equation}\label{16}
    h^i_{\mu} (\nabla_{\nu} T^{\mu\nu})=(\rho+p)u^{\nu}(\nabla_{\nu}u^{i})+(g^{i\nu}+u^i u^{\nu})(\partial_{\nu}p)=0 ,
\end{equation}
and a scalar equation as,
\begin{equation}\label{5}
    u_{\mu} (\nabla_{\nu} T^{\mu\nu})=u^{\mu}\left(h \partial_{\mu}\tilde{\rho}-\partial_{\mu}\rho\right)=0.
\end{equation}
In equation (\ref{16}), the roman index assumes $i=1,2,3$. 
Conventionally, the evaluation of the accreting matter is described in terms of the quantity defined in co-rotating frame of the fluid considering the following series of variable transformations: Azimuthal velocity is defined as $v_{\phi}^2=(u^{\phi}u_{\phi})/(-u^t u_t)$ with the associated Lorentz factor $\gamma_{\phi}=1/\sqrt{(1-v^2_{\phi})}$; the polar velocity is defined as $v_{\theta}^2=\gamma^2_{\phi}(u^{\theta}u_{\theta})/(-u^t u_t)$ with the associated Lorentz factor $\gamma_{\theta}=1/\sqrt{(1-v^2_{\theta})}$, and finally the radial velocity is defined as $v^2=\gamma^2_{\phi}\gamma^2_{\theta}(u^r u_r)/(-u^t u_t)$ with the associated radial Lorentz factor $\gamma_v=1/\sqrt{(1-v^2)}$. All detailed equations formulated using the co-rotating fluid variables are provided in the reference \cite{Dihingia:2018tlr}.
% All the detailed equations in terms of those co-rotating fluid variables can be found in the reference \cite{Dihingia:2018tlr}. 
However, for our present study we focus on spherical accretion and assume $v_{\theta}=v_{\phi}=0$~\cite{1984ApJ...277..296H}. Expanding the Eq.~(\ref{16}) for $i=r$, we get,
\begin{equation}\label{18}
    v \gamma^2_v \frac{dv}{dr}+\frac{1}{h\tilde{\rho}}\frac{dp}{dr}+\frac{d\Phi^{\text{eff}}}{dr}=0 .
\end{equation}
% {\red Is this sentence needed?$\rightarrow$ Note the relativistic Lorentz factor $\gamma_v$ and the general relativistic effective potential $\Phi^{\text{eff}} =\frac{1}{2}\ln (-g_{tt})$, which were not taken into account in earlier studies $\leftarrow$.} {\blue agreed} 
Here, $\Phi^{\text{eff}} ~[ =\frac{1}{2}\ln (-g_{tt})]$ is the general relativistic effective potential and $h ~[=\rho+p)/\tilde{\rho}]$ denotes the enthalpy of the flow.
%{\red Note that the relativistic Lorentz factor $\gamma_v$ and the general relativistic effective potential $\Phi^{\text{eff}}$ were not considered in earlier studies~\cite{Ricotti:2007jk,Wang:2023qxj,Ali-Haimoud:2016mbv}.} \textcolor{blue}{We can remove this}.
% is $h = (\rho+p)/\tilde{\rho}$.  %Simillarly from equation (\ref{17}) we get
%\begin{equation}\label{20}
%    h\frac{d\tilde{\rho}}{d r}-\frac{d \rho}{dr}=0.
%\end{equation}
%In the next section we consider a certain BH geometry to obtain the dynamical equations.
%\subsection{Evolution equations in Schwarzschild geometry}
% For our present study 
In this work, we consider Schwarzschild BH where $g_{tt}= -g^{rr} =- f(r),~g_{t\phi} =0,~g_{\theta\theta} =g_{\phi\phi}/\sin^2 \theta = r^2$ with $f(r)=1-{2 G M}/{r}$. Here, $M$ denotes the mass of the BH. The effective gravitational potential perceived by the fluid is given by $\Phi^{\text{eff}} =\frac{1}{2}\ln f(r)$, and the radial momentum equation transforms into,
\begin{equation}\label{radial_eq}
    v \gamma^2_v \frac{dv}{dr}+\frac{1}{h\tilde{\rho}}\frac{dp}{dr}+ \frac 1 2 \frac{f'(r)}{f(r)} =0.
\end{equation}
The mass accretion rate is obtained from the continuity as,
\begin{equation}\label{23}
    \frac{dM}{dt}=4\pi r^2 \tilde{\rho}\gamma_v v\sqrt{f(r)}.
\end{equation}
Together, Eqs.~\eqref{radial_eq} and~\eqref{23} constitute the fundamental set of equations that describe the dynamics of steady-state, spherical accretion onto a BH in an asymptotically flat spacetime.

In the non-relativistic (Newtonian) limit, valid at large radial distances from the source, the fluid velocity is sub-luminal, leading to $\gamma_v \to 1$ and $h \to 1$. Under these approximations, the general relativistic equations of motion reduce to their well-known Newtonian forms
\begin{equation}\label{v_equation_non_rel}
    v \frac{dv}{dr}+\frac{1}{\rho}\frac{dp}{dr}+  \frac{GM}{r^2} =0,
\end{equation}
and the mass accretion rate becomes
\begin{equation}\label{Acc_eq_non_rel}
    \frac{dM}{dt}=4\pi r^2 \rho v.
\end{equation}
Most of the studies of accretion onto PBHs have employed these Newtonian equations to model the process within a cosmological context~\cite{Ali-Haimoud:2016mbv,Ricotti:2007jk,Wang:2023qxj,Tabasi:2022fap}. We will show how relativistic corrections in Eqs.~\eqref{radial_eq} and \eqref{23} change the outcome compared to the Newtonian one.

We have obtained the general relativistic Hydrodynamic equation in static black hole background. In the following we intend to apply this in the cosmological context where background spacetime is expanding. Exact analysis for such generalization would be to find out the black hole solution in cosmological background which is beyond the scope of our present work. To  incorporate the expansion of the universe directly we consider the Friedmann–Lemaître–Robertson–Walker (FLRW) background, parametrized by the scale factor $a(t)$, with the metric $ds^2 = -dt^2 + a^{2}(t)(dr^2 + r^2 d\theta^2 + r^2 \sin^2 \theta\, d\phi^2)$, and follow the approximation detailed in \cite{Ricotti:2007jk}.

\subsection{Accretion in the expanding background}

The background expansion is incorporated by redefining the radial coordinate $r = a(t) x$. In this expanding coordinate, the radial velocity $v$ in terms of comoving coordinate $x$ and peculiar velocity $v_p$ becomes,
\begin{equation}
    v =\frac{dr}{dt}=Hax+v_p ,\label{v_p}
\end{equation}
where $ H~(=\dot{a}/a)$ is the Hubble parameter and $v_p~(=a \dot{x})$. The Hubble parameter $H$ satisfies the dynamical equation $H^2=\dot{a^2}/a^2 = ({1}/{3M_p^2}) \rho$. Where, $\rho$ corresponds to the background energy density of the Universe.

In all practical purpose the BH horizon radius $(r_s)$ satisfies $r_s \ll H^{-1} \rightarrow \epsilon = r_s H \ll 1$. Therefore, asymptotic flat black hole approximation which we considered in the earlier discussion can be approximately valid. One can further define a length scale, $r_*$ associated with the sphere of influence of a BH within which its gravity tends to dominate over the cosmic expansion. It can be approximately defined as the total background mass  contained in the volume of radius $r_*$ equals the BH mass, as $M = \frac{4}{3}\pi r_*^3 \rho$. This yields $r_* = r_s/\epsilon^{2/3}$. Thus, for small $\epsilon$, the gravitational influence extends far beyond the Schwarzschild radius ($r_* \gg r_s$). Therefore, Within this region, the fluid dynamics are governed by the quasi-static Schwarzschild geometry with Hubble flow being treated as small perturbation to the accretion flow in this limit. However $r>r_*$, background can be described by standard cosmological expansion with the scale factor $a(t)$.   

%Our formalism correctly captures this physical reality, a
%Therefore,  in the region of our interest $r> r_s$ is validity of this method in two relevant physical limits:
%\begin{itemize}
%\item \textbf{Near the Black Hole $(r \ll H^{-1})$:} In this inner region, the gravitational potential of the black hole strongly dominates over the cosmic expansion. The fluid dynamics are governed almost entirely by the quasi-static Schwarzschild geometry. Our formalism correctly captures this physical reality, as the Hubble flow becomes a small perturbation to the accretion flow in this limit.
%\item \textbf{Far from the Black Hole $(r \sim H^{-1})$:} At large distances, the fluid is expected to asymptotically follow the Hubble flow. Our substitution $r(t)=a(t)x$ naturally recovers this cosmological behavior by construction, ensuring the correct large-scale kinematics.
%\end{itemize}
%This approach effectively serves as a physical interpolation between the two regimes: a gravity-dominated, quasi-static flow near the black hole and a Hubble-dominated flow at cosmological distances.

\begin{figure}
    \centering
    \includegraphics[width=\columnwidth]{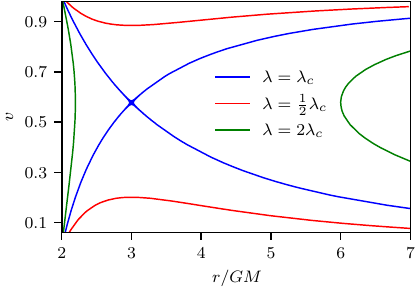}
    \caption{Plot of flow velocity ($v$) as a function of radial distance ($r$) for different accretion rate, with an equation of state 
    % considering an equation of state with 
    $\omega=1/3$.}
    \label{critical}
\end{figure}

Incorporating Eq. (\ref{v_p}) into the accretion equations, and integrating Eq.~(\ref{radial_eq}) we get,
\begin{align}\label{29}
    \ln \left(1-(Hax+v_p)^2\right)+\frac{2\omega}{(1+\omega)}\ln \left(\frac{\rho_{\infty}}{\rho} \right)
    -\ln f(ax)=0.
\end{align}
The integration constant is set by the condition that far away from the BH ($x \rightarrow \infty$), $v \rightarrow 0$, and $\rho \rightarrow \rho_\infty \propto a^{-3(1+\omega)}$, which corresponds to the background energy density of the Universe, characterized by the generic isothermal equation of state $p = \omega \rho$.
% The integration constant is set by the condition that at far away from the BH {\orange ($x \rightarrow \infty$)}, $v \rightarrow 0$, and $\rho \rightarrow \rho_\infty \propto a^{-3(1+\omega)}$ that is the background energy density of the Universe, that is characterize by the generic isothermal equation of state $p = \omega \rho$. 
The Hubble parameter $H^2=\dot{a^2}/a^2 = ({1}/{3M_p^2}) \rho$, can be further expressed as $H=\Hin  \left(a/\ain\right)^{-\frac{3}{2}(1+\omega)}$, with index ``${\rm in}$" being some initial time which we identify later. Here, $M_p=1/\sqrt{8\pi G} \simeq 2.435\times 10^{18}\text{Gev}$, being the reduced Planck mass. Utilizing this, the relation between energy density and mass density of the fluid can be obtained by integrating the enthalpy generation Eq.~(\ref{5}) as,
\begin{equation}\label{30}
\tilde{\rho}^{1+\omega}=\rho_{\infty}^{\omega}\rho.
\end{equation}
To obtain this, we have assumed the mass and energy densities of the fluid at infinity are equal to $\rho_{\infty}$. Further, using Eq.~(\ref{v_p}) and Eq.~(\ref{30}), Eq.~(\ref{23}) can be written as,
\bea
\frac{dM}{dt}=4\pi (ax)^2 \rho_{\infty}^{\frac{\omega}{1+\omega}}\rho^{\frac{1}{1+\omega}} \frac{(Hax+v_p)}{\sqrt{1-(Hax+v_p)^2}} \sqrt{f(ax)} .\;\; \label{29_a}
\eea
The mass accretion rate $\dot{M}~(=dM/dt)$ is assumed to be a conserved quantity, and satisfies $d\dot{M}/dx=0$. Utilizing this and combining Eq. (\ref{29}) and (\ref{29_a}), one readily obtains the radial variation of the comoving fluid velocity as, 
\begin{equation}
    \frac{dv_p}{dx}=\frac{\frac{2\omega}{(ax)}  + \frac{\omega f'(ax)}{2 f(ax)} - \frac{f'(ax)}{2 f(ax)} 
-   H \frac{v^2 -\omega}{v(1-v^2)}}{\frac{v^2 -\omega}{v(1-v^2)} \frac{1}{a}} =\frac{N}{D}.
\end{equation}
In Fig.~\ref{critical}, we have depicted the various representative
% plotted all possible 
solution topologies for the inflowing matter. The physical solutions are the ones where the inflowing matters pass through a critical point and transits smoothly from subsonic to supersonic branch. At the critical point, 
% For this 
one needs to satisfy $N = D = 0$ condition simultaneously. Solving this, we get the critical point location as $x_c= GM (1+3\omega)/(2 a \omega)$ and matter velocity at $x_c$ as $v^c_p=\sqrt{\omega}-Hax_c$. Inserting these in Eq.~(\ref{29_a}), the accretion rate at the critical point becomes,
%%%%%%%%%%%%%%%%%%%%%%%%%%%%%%%%%%%%
%As matter is falling radially into the black hole, the flow must undergo from sub-soninc to super-sonic flow passing though the critical point. 
%Using (\ref{29}), it can further simplify to
%\begin{align}\label{32}
% f(ax)^{\frac{1-\omega}{2\omega}}  \frac{1}{(ax)^2} \frac{\dot{M}
%}{4 \pi \rho_{\infty}}= & (Hax+v_p)  %\notag \\
%& \times \left(1-(Hax+v_p)^2\right)^{\frac{1-\omega}{2\omega}}
%\end{align}
%The right hand side of the above equation has a minima at $v_p=\sqrt{\omega}-Hax$ and left hand side has a minima at $x = GM (1+3\omega)/(2 a \omega)$. These two are also called critical points. Putting those values in equation (\ref{32}) we get the mass accretion rate
%\begin{equation}\label{31}
 %   \dot{M}=4\pi \left( \frac{1}{4}\omega^{-\frac{3}{2}} \left(1+3\omega\right)^{\frac{1+3\omega}{2\omega}} \right) G^2 M^2 \rho_{\infty}
%\end{equation}
\begin{equation}\label{10}
     \frac{dM}{dt}=\frac{\lambda_c}{16 \pi M_p^4}  M^2 \rho_{\infty},
     %4 \pi \lambda_c G^2 M^2 \rho_{\infty},
\end{equation}
where $\lambda_c=(1/4)\omega^{-3/2}(1+3\omega)^{(1+3\omega)/2\omega} $. Following the same procedure for the non-relativistic equations, namely Eq.~\eqref{v_equation_non_rel} and Eq.~\eqref{Acc_eq_non_rel}, the location of the critical point is determined to be $x_c^{\rm NR} = GM/(2a\omega)$. At this point, the peculiar velocity of the matter is given by $v_{p,c}^{\rm NR} = \sqrt{\omega} - GMH/(2\omega)$. These values, in turn, define the accretion rate expressed in Eq.~\eqref{10}, with the non-relativistic dimensionless critical eigenvalue $\lambda_c^{\rm NR}$ taks the form $\lambda_c^{\rm NR}=(1/4)\exp(3/2) \omega^{-3/2}$, which is consistent with the results presented in~\cite{Ali-Haimoud:2016mbv} for an isothermal equation of state.
%Note that in the non-relativistic limit, the critical value reduces to $\lambda_c^{\rm NR}=(1/4)\text{exp}(3/2) \omega^{-3/2}$~\cite{Ali-Haimoud:2016mbv}. 
Integrating Eq.~(\ref{10}), we obtain
\bea 
\frac{M}{\Min} = \left[ 1-\frac{\lambda_c \gamma}{2(1+\omega)}\left( \frac{\frac{3}{2}\Hin(1+\omega)(t-\tin)}{\frac{3}{2}\Hin(1+\omega)(t-\tin)+1} \right) \right]^{-1} \;\label{acc_mass_evolution}
\eea
where $\Min~(=4 \pi \gamma M_p^2 \Hin^{-1})$ is the initial mass of a PBH with $\Hin$ being the Hubble parameter at the formation time $t=\tin$. Here, $\gamma~(=\omega^{3/2})$ is the collapse fraction to form PBH \cite{1974MNRAS.168..399C}. 
%Here we have used the equation of state of the background fluid as $p_{\infty}=\omega \rho_{\infty}$. 
From Eq.~(\ref{acc_mass_evolution}), we indeed see that PBH gains mass due to accretion and quickly saturates into (see Fig.~\ref{Mass_evo}), 
\begin{equation}\label{17}
M_{\rm acc}= \Min\left(1-\frac{\lambda_c \gamma}{2(1+\omega)}\right)^{-1}.
\end{equation}
This is one of the key results of our analysis. For example, setting $\omega = 1/3$ leads to a striking increase in the BH mass, reaching approximately $4.53\, \Min$\footnote{Throughout our analysis, we assume non-rotating PBHs. However, if spinning PBHs are considered, the mass increase would be slightly higher—a possibility we leave for future investigation.}
% Note that throughout our analysis, we assume non-rotating PBHs. If we instead consider spinning PBHs, the mass increment turns out to be slightly higher a possibility we leave for future analysis.}. 
% For example, if we consider $\omega=1/3$, the BH mass enhance to $\simeq 4.53 M_i$
The accretion time scale of attaining this mass can immediately be obtained from Eq.~(\ref{acc_mass_evolution})  at the time when accretion rate equals expansion rate, $i.e.$, when $\dot{M}/M = H$, which yields,
% and expansion rate becomes equal $\dot{M}/M = H$ which yields, 
%\textcolor{red}{
%\begin{align*}
%    t_{\rm acc} - t_{\rm i} = 
    %\tau_{\rm BH}^{\rm acc} = & \frac{3}{4\pi\gamma}\frac{M_i}{M_p^2} \frac{2}{9(1+\omega)}\\
%    & \times \left\lbrace \frac{3\gamma \lambda_c}{4}\left(1- \frac{\lambda_c\gamma}{2(1+\omega)} \right)^{-1} -1 \right\rbrace \\
%    \simeq \frac{3}{4\pi\gamma}\frac{M_i}{M_p^2} \times 1.08
%\end{align*}}
%%%%%%%%%%%%%%%%%%%%%%%%%%%%%%%%%%%%%%%%%%%%
\begin{equation}
  t_{\rm acc} - \tin = \tau_{\rm BH}^{\rm acc} \simeq \frac{3.24}{4\pi\gamma}\frac{M_{\rm in}}{M_p^2}.
\end{equation}
For example, if we consider a PBH of mass $10$ g accreting in the radiation background ($\omega =1/3$), accretion ceases at approximately $ \tau_{\rm BH}^{\rm acc} \simeq 3 \times 10^{6} M_p^{-1}$.

%$ \tau_{\rm BH}^{\rm acc} \simeq 4.7 \times 10^{5} M_p^{-1}$

%\subsection{Verification of the critical points using $dv_p/dx$ procedure}
%The mass accretion rate $\dot{M}$ is independent of position from the BH center. So, from the condition $d\dot{M}/dx=0$, we can derive an expression for $d\rho/dx$ as
%\begin{align}
 %   \frac{d \rho}{dx}=-(1+\omega)\rho \Bigg\lbrace \frac{2}{x}+ \frac{1}{(Hax+v_p)} \frac{1}{(1-(Hax+v_p)^2)} \notag \\ \frac{d(Hax+v_p)}{dx}  +  \frac{1}{\left(1-\frac{2GM}{(ax)}  \right)} \frac{aGM}{(ax)^2} \Bigg\rbrace .
%\end{align}
%Now we insert this into the radial equation (\ref{22}) and solve for $dv_p/dx$ as
%\begin{equation}
 %   \frac{dv_p}{dx}=\frac{N}{D}=\frac{\frac{2\omega}{(ax)}  + \frac{\omega f'(ax)}{2 f(ax)} - \frac{f'(ax)}{2 f(ax)} 
%-   H \frac{v^2 -\omega}{v(1-v^2)}}{\frac{v^2 -\omega}{v(1-v^2)} \frac{1}{a}}
%\end{equation}
%where
%\begin{align}
%N =  & \frac{2\omega}{(ax)}  + \frac{\omega f'(ax)}{2 f(ax)} - \frac{f'(ax)}{2 f(ax)} 
%-  a D 
%\frac{(v_p+Hax)^2 -\omega}{(v_p+Hax)(1-(v_p+Hax)^2)}
%H ,
%\end{align}
%and
%\begin{align}
%D = & \frac{v^2 -\omega}{v(1-v^2)} \frac{1}{a} .
%\end{align}
%Now the sonic points of $x$ and $v_p$ are where $N$ and $D$ are both simultaniously zero. So, $D=0$ gives $v_p=\sqrt{\omega}-Hax$ and $N=0$ gives $x = GM (1+3\omega)/(2 a \omega)$. So, this procedure also gives us the same critical points. Inserting those values of $x$ and $v_p$ into the accretion rate equation (\ref{29_a}), gives equation (\ref{31}).

\section{Hawking evaporation}

In the quantum mechanical framework, a BH can also emit particles via Hawking radiation.
%The particle emission rate per unit momentum interval from a Schwarzschild BH is expressed as,
%\begin{equation}\label{37}
%    \frac{d^2 N_j}{dp dt}=\frac{g_j}{2\pi^2}\frac{\sigma_{s_i}(E_j)}{e^{E_i/T_{\rm BH}}- (-1)^{2s_j}} \frac{p^3}{E_j}
%\end{equation}
%\begin{equation}
%n_{E_j} =\frac{\Gamma_{s_j}(E_j)}{e^{E_j/T_{\rm BH}}- (-1)^{2s_j}}, 
%\end{equation}
%where $T_{\rm BH}=M_p^2/M $ is the BH temperature, $M_p=1/\sqrt{8\pi G} \simeq 2.435\times 10^{18}\text{Gev}$ being the reduced Planck mass, $\Gamma_{s_j}(E_j)$ is the Graybody factor and $s_j$ is the spin of the $j^{\text{th}}$ species of mass $m_i$ and momentum $p$ having energy $E_j^2 =m^2_j+p^2$. %Utilizing these the particle emission rate per momentum interval becomes, \begin{equation}\label{37}
%    \frac{d^2 N_j}{dp dt}=\frac{g_j}{2\pi^2}\frac{\sigma_{s_i}(E_j)}{e^{E_i/T_{\rm BH}}- (-1)^{2s_j}} \frac{p^3}{E_j}
%\end{equation}
%where $g_j$ is the internal degrees of freedom of the emitted particles and $\sigma_{s_j}(E_j)$ is the scattering crossection of the BH related to the Graybody as $\Gamma_{s_j}(E_j)=\sigma_{s_j}(E_j) p^2/\pi$. 
Due to this Hawking emission, BH loses its mass in the cosmological time scale.
%and its decay rate is expressed as,
%the cross section assumes $\sigma_{s_j}(E_j) \simeq 27 M^2/64\pi M_p^4$. From this we now read two important quantities of our interest. 
%Due to this Hawking emission BH loses its mass in the cosmological time scale, 
The BHs decay rate can be well described by~\cite{Page:1976df},
\begin{equation}\label{22}
    \frac{dM}{dt} 
    %=-\sum_{j}\int_0^{\infty}E_j \frac{d^2 N_j}{dp dt}dp
    =-\epsilon(M) \frac{M^4_{p}}{M^2}.
\end{equation}
In the high energy limit of the emitted particles $E_j >> T_{\rm BH}$, called geometrical optical limit, the parameter $\epsilon(M)$ %is very weakly dependent on mass and 
is approximately expressed as $\epsilon \simeq \frac{27}{4} \frac{g_{\ast}(T_{\rm BH}) \pi}{480}$, with $ g_{\ast}(T_{\rm BH}) \sim 106.76$ being the number of relativistic degrees of freedom at the BH temperature $T_{\rm BH}=M_p^2/M$ \cite{Baldes:2020nuv}.
%where $\epsilon(M)=\sum_{j}g_j \epsilon_j(z_j)$ and 
%\begin{equation}
%\epsilon_j(z_j)=\frac{27}{64 \pi^3}\tilde{\epsilon} e^{z_j} \left[ z_j^2 Li_2(\tilde{\epsilon})+3z_j Li_3(\tilde{\epsilon})+3 Li_4(\tilde{\epsilon})\right].
%\end{equation}
% \begin{equation}
% \epsilon_i(z_i)= \frac{27}{128 \pi^3}\int_{z_i}^{\infty}\frac{(x^2-z^2_i)}{e^{x}- (-1)^{2s_i}} x dx.
% \end{equation}
%Here $\tilde{\epsilon}=(-1)^{2s_i} e^{-z_i}$, $z_i=m_i/T_{\rm BH}$ and $Li_n$ is the Polylog function of order $n$ \cite{jhg}. 
With this assumption, the PBH mass evolves as,
 %%%%%%%%%%%%%
\begin{equation}
M \simeq \Min\left[1-\frac{(t-\tin)}{\tau_{\rm BH}^{\rm ev}}\right]^{\frac{1}{3}}.
\end{equation}
Similar to the accretion time scale, here also we have BH evaporation time scale set by $M=0$,
\begin{eqnarray}\label{lifetime}
t_{\rm ev} - \tin= \tau_{\rm BH}^{\rm ev} \simeq \frac{M_{\rm in}^3}{3 \epsilon M_p^4}.
\end{eqnarray}
For example, assuming a PBH with a mass of 10 g formed in a radiation-dominated background ($\omega = 1/3$), it completely evaporates at approximately $\tau_{\rm BH}^{\rm ev} \simeq 8.8 \times 10^{17} M_p^{-1}$. This highlights the wide separation between the timescales of the two different physical processes of our interest, namely accretion and evaporation ($\tau_{\rm BH}^{\rm ev} \gg \tau_{\rm BH}^{\rm acc}$), as depicted in Fig.~\ref{Mass_evo}.

% For example, assuming
% %$\epsilon = \pi g_{*}/{480}$, with $ g_{*}\sim 106$ the number of degrees of freedom below $T_{\rm BH }$ \cite{Masina:2020xhk}), and 
% a PBH of mass $10$ g emerged in radiation bath ($\omega =1/3$), completely evaporates at around $ \tau_{\rm BH}^{\rm ev} \simeq 8.8 \times 10^{17} M_p^{-1}$. Note the wide separation of time scales of two different physical processes of our interest namely accretion and evaporation ($\tau_{\rm BH}^{\rm ev} \gg \tau_{\rm BH}^{\rm acc}$) as depicted again in the left panel of Fig. \ref{accretion}.

%\cite{Masina:2020xhk}

The second natural quantity of our interest is the number of particles produced during its entire life time of a PBH. 
%particle production rate $\Gamma_{\rm BH \rightarrow i}$ for individual species, 
%and the total number of particles produced from the BH $N_i$. Integrating equation (\ref{37}) over momentum we get
%\begin{equation}\label{25}
 %   \Gamma_{\rm BH \rightarrow i} =\frac{dN_i}{dt}=\int_0^{\infty}dp\frac{d^2N_i}{dp dt}=g_i \psi_i(z_i) \frac{M_p^2}{M} 
%\end{equation}
%where 
%\begin{equation}
%$\psi_i(z_i) 
%=\frac{27}{64 \pi^3} \tilde{\epsilon} \left[ z_i Li_2(\tilde{\epsilon}e^{-z_i})+ Li_3(\tilde{\epsilon}e^{-z_i})\right] 
%\simeq \frac{27}{64 \pi^3}\zeta(3)$.
%\end{equation}
% \begin{equation}
% \psi_i(z_i)=\frac{27}{128 \pi^3} \int_{z_i}^{\infty} dx \frac{(x^2-z_i^2)}{e^{x}- (-1)^{2s_i}} .
% \end{equation}
%Therefore, the total number of particles produced can be
%obtained by integrating over its lifetime $\tau$ as $N_i =\int_{0}^{\tau}\Gamma_{\rm BH \rightarrow i}dt$. 
%for two different situations. 
Following \cite{Haque:2023awl,Haque:2024eyh,Barman:2024slw}, one can deduced the total number of emitted particles by a PBH of modified mass $M_{\rm acc}$, and temperature $T_{\rm BH}^{\rm acc} =T_{\rm BH}^{\rm in} \left( 1-\frac{3 \lambda_c \gamma}{8} \right)$ due to accretion as,
%If mass of an emitted particle is $m_i \lesssim T_{\rm BH}^{\rm acc} = T_{\rm BH}^{\rm i} \left( 1-\frac{3 \lambda_c \gamma}{8} \right)$ with $ T_{\rm BH}^{\rm i} = {M_p^2}/{M_i}$ being the initial BH temperature, 
%one can consider the particle production is effective throughout the BH lifetime. 
\begin{equation}\label{N_dm}
    N_j
%    = \int\int_{t_i}^{t_{\rm ev}} \frac{d^2N_j}{dt dp}dtdp 
    %\simeq \frac{15 g_i \zeta(3)}{g_{\ast}\pi^4} \frac{M_{\rm acc}^2}{M_p^2}
    \simeq 
    \begin{cases}
       10^8 \left( \frac{M_{\rm acc}}{1 \; \rm g} \right)^2 ~~~~~ \text{for}~ m_j\lesssim T_{\rm BH}^{\rm acc}, \\
       10^{14} \left( \frac{10^{10}\text{Gev}}{m_j} \right)^2  ~\text{for}~ m_j\gtrsim T_{\rm BH}^{\rm acc}, \\
       %(k/k_{\rm end})^{-2\nu_2} \quad &\text{for} \quad \xi>3/16 
        \end{cases}     
%    10^8 \left( \frac{M_{\rm acc}}{1 \; \rm g} \right)^2 ,
\end{equation}
%and for $m_i \gtrsim T_{\rm BH}^{\rm acc} $
%, then we need to integrate the equation (\ref{25}) from $\tilde{t}_i$ to $t_{\rm ev}$, where $\tilde{t}$ is the time where $m_i=T_{\rm BH}$. Simple calculation gives
%\begin{equation}
%    \tilde{t}=\tau_{\rm BH}^{\rm ev}\left(1-\frac{M_p^6}{m_i^3 M_i^3} \right),
%\end{equation}
%the total number of emitted particles will be
%\begin{equation}\label{N_dm_b}
 %   N_i^{m_i > T_{\rm BH}^{\rm acc}}
    %= \int_{\tilde{t}}^{t_{ev}} \frac{dN_i}{dt}dt \simeq \frac{15 g_i \zeta(3)}{g_{\ast}\pi^4} \frac{M_p^2}{m_i^2}
%    \simeq 10^{14} \left( \frac{10^{10}\text{Gev}}{m_i} \right)^2 .
%\end{equation}
where `$j$' stands for a particular emitted particle of mass $m_j$, and $T_{\rm BH}^{\rm in}~(=M_p^2/M_{\rm in})$ is the initial BH temperature. Note that Eq.~(\ref{N_dm}) considers only scalar particles; for fermionic particles, a factor of $3/4$ should be included.

\section{PBH mass evolution}

In the preceding sections, we have discussed two important physical processes under which PBH evolves. Besides accretion being responsible for its increase in mass, a PBH also loses its mass by emitting Hawking radiation. Both the processes is described by the following PBH mass evolution equation as,
\begin{equation}
    \frac{dM}{dt}= \left( \frac{dM}{dt} \right)_{\text{accretion}}+\left( \frac{dM}{dt} \right)_{\text{evaporation}}.
    \label{accev}
\end{equation}
We note that while accretion takes place early in the PBH's lifetime over a very short timescale $\tau_{\rm BH}^{\rm acc}$, evaporation occurs much later, near the end of its life.
% We noticed that whereas accretion occurs near the initial stage of PBH's life within a very short time scale $\tau_{\rm BH}^{\rm acc}$, the evaporation, on the other hand, occurs near its end of life. 
Due to this widely separated time scale of processes, we analytically solve Eq.~(\ref{accev}) in two steps. 
As delineated earlier, in the first step, accretion enables the PBH to quickly gain mass, reaching $M_{\rm acc} = \Min \left(1 - \frac{\lambda_c \gamma}{2(1 + \omega)}\right)^{-1}$. Accretion ceases at time $t_{\rm acc} - \tin  = \tau_{\rm BH}^{\rm acc}$, when the accretion rate equals the cosmic expansion rate, thereby setting the initial condition for the PBH's subsequent evaporation phase.
% As delineated earlier, in the first step the accretion helps the PBH to quickly gain its mass say $M_{\rm acc} = M_{i} ( 1-{\lambda_c \gamma}/{2(1+\omega)})^{-1}$. At time $t_{\rm acc}-t_{i} = \tau_{\rm BH}^{\rm acc}$, accretion stops once its rate becomes equal to the expansion rate, and sets the initial condition for the subsequent evaporation stage the PBH. 
Upon solving the evaporation equation, we find that
\begin{equation}\label{evolution_eq}
M \simeq M_{\rm in}\left[1-\frac{\lambda_c \gamma}{2(1+\omega)}\right]^{-1}\left[1-\frac{(t-t_{\rm acc})}{\tau_{\rm BH}}\right]^{\frac 1 3},
\end{equation}
where the PBH life time modifies into the following approximate form as,
\begin{equation}\label{PBH_lifetime}
\tau_{\rm BH} \simeq \frac{\Min^3}{3 \epsilon M_p^4}\left[1-\frac{\lambda_c \gamma}{2(1+\omega)}\right]^{-3}.
\end{equation}

\begin{figure}[t]
    \centering    \includegraphics[width=0.9\linewidth]{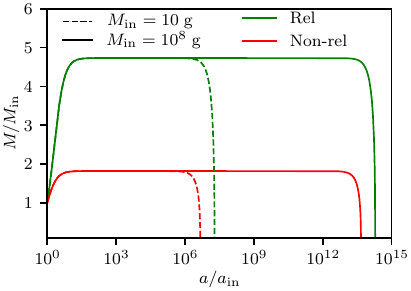}
    \caption{PBH mass ($M$) evolution as a function of scale factor ($a$) for initial masses of $10$ g (dashed) and $10^8$ g (solid), comparing relativistic (green) and non-relativistic (red) accretion along with Hawking evaporation.}
    \label{Mass_evo}
\end{figure}

The evolution of PBH mass, accounting for both relativistic (blue) and non-relativistic (red) accretion in conjunction with Hawking evaporation, is illustrated in Fig.~\ref{Mass_evo}. Results are presented for two initial masses, namely $10\, \mathrm{g}$ (dashed lines) and $10^8\, \mathrm{g}$ (solid lines), respectively.

\section{BBN Constraint}

PBHs evaporating during the era of Big Bang Nucleosynthesis (BBN) can inject additional relativistic degrees of freedom, potentially altering the primordial element abundances predicted by BBN. To preserve the observed consistency of BBN results, PBHs must evaporate before the onset of BBN. This requirement immediately suggests that the PBH mass within the range of $10^8 \,\text{g} \lesssim M_{\rm BH} \lesssim 10^{13} \,\text{g}$
% $M_{\rm BH} \sim 10^8 \,\text{g}$ to $10^{13} \,\text{g}$ 
\cite{Carr:2020gox,Keith:2020jww,Carr:2009jm} should not form in the early universe.  However, when PBH accretion is taken into account, these mass constraints are naturally altered.
% However, the inclusion of PBH accretion naturally modifies these restriction of mass range.

%From the PBH mass evolution Eq.~(\ref{evolution_eq}), one computes the evaporation temperature by setting $M=0$ yielding the following relation, 

\begin{figure}[t]
    \centering
    \includegraphics[width=0.9\linewidth]{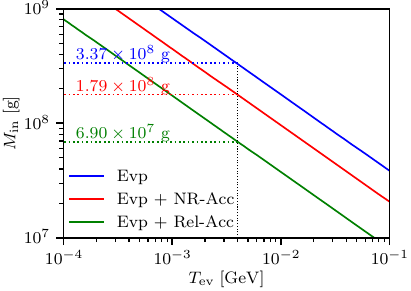}
    \caption{Initial PBH mass ($M_{\rm in}$) as a function of evaporation temperature ($T_{\rm ev}$), shown for three cases namely, blue, red, and green curves representing without, with non-relativistic, and with relativistic accretion respectively.}
    \label{BBN_plot}
\end{figure}

In order to calculate the evaporation temperature ($T_{\rm ev}$), we set $\omega =1/3$ and rewrite the PBH lifetime (Eq. \eqref{PBH_lifetime}) as,
\begin{equation}\label{24}
    \tau_{\rm BH}    %t_{\rm ev}-t_{\rm acc}=
   \simeq\frac{1}{2H_{\rm acc}} \left(\frac{a_{\rm ev}}{a_{\rm acc}}\right)^{2} = \frac{\Min^3}{3 \epsilon M_p^4}\left[1-\frac{3\lambda_c \gamma}{8}\right]^{-3},
     \end{equation}
where $H_{\rm acc}~(= 4\pi \gamma M_p^2 M_{\rm acc}^{-1})$ is the Hubble parameter at the end of accretion $t_{\rm acc}$. In this radiation dominated Universe with energy density $\rho_r=\frac{\pi^2}{30} g_{\ast}(T) T^4$, the scale factor and temperature are related as $a_{\rm ev}/a_{\rm acc}=T_{\rm acc}/T_{\rm ev}$, and hence, PBH evaporation temperature ($T_{\rm ev}$) is computed as,
\bea
T_{\rm ev} \simeq \sqrt{\frac{3\epsilon}{2}} \left( \frac{\pi^2 g_{\ast}}{30} \right)^{-\frac{1}{4}} \left( \frac{M_p}{\Min} \right)^{\frac{3}{2}} \left(1-\frac{3\gamma\lambda_c}{8} \right)^{\frac{3}{2}} M_p ,~~ \label{25}
\eea
where the relation \( T_{\rm acc} =(\pi^2 g_{\ast}/90)^{-1/4} (H_{\rm acc} M_p)^{1/2} \) is used corresponding to radiation temperature at $H_{\rm acc}$.
% The evaporation temperature can be computed utilizing the expression of PBH lifetime {\orange Eq.} \eqref{PBH_lifetime} in the following manner,
% \begin{equation}
%     \tau_{\rm BH}    %t_{\rm ev}-t_{\rm acc}=
%    \simeq\frac{1}{2H_{\rm acc}} \left(\frac{a_{\rm ev}}{a_{\rm acc}}\right)^{2} = \frac{M_{i}^3}{3 \epsilon M_p^4}\left[1-\frac{3\lambda_c \gamma}{8}\right]^{-3},
%      \end{equation}
% where $H_{\rm acc}= 4\pi \gamma M_p^2 M_{\rm acc}^{-1}$ is the Hubble parameter at the end of accretion $t_{\rm acc}$. In the above we set $\omega =1/3$, and in this radiation dominated Universe with energy density $\rho_r=\frac{\pi^2}{30} g_{\ast}(T) T^4$, the scale factor and temperature are related as $a_{\rm ev}/a_{\rm acc}=T_{\rm acc}/T_{\rm ev}$, and hence PBH evaporation temperature $T_{\rm ev}$ becomes,
% \begin{equation}
%     T_{\rm ev} \simeq \sqrt{\frac{3\epsilon}{2}} \left( \frac{\pi^2 g_{\ast}}{30} \right)^{-1/4} \left( \frac{M_p}{M_i} \right)^{3/2} \left(1-\frac{3\gamma\lambda_c}{8} \right)^{3/2} M_p ,
% \end{equation}
% where, the relation \( T_{\rm acc} =(\pi^2 g_{\ast}/90)^{-1/4} (H_{\rm acc} M_p)^{1/2} \) has been used corresponding radiation temperature at $H_{\rm acc}$.
% formation, related to the initial PBH mass \( M_i \) by:
% \begin{equation}\label{initial_temp}
% T_i = M_i \left( 4 \pi \gamma \right)^{1/2} \left( \frac{\pi^2}{90} g_{\ast}(T_i) \right)^{-1/4} \left( \frac{M_p}{M_i} \right)^{3/2}.
% \end{equation}
% $T_{\rm acc}\simeq T_i $ the initial radiation temperature from equation (\ref{initial_temp}) replacing $M_i$ by $M_{\rm acc}$.
Combining Eqs.~(\ref{24}) and (\ref{25}), the initial PBH mass can be expressed as a function of the evaporation temperature,
% Combining these relations, Eq.~(\ref{24}) and (\ref{25}), the initial PBH mass is expressed as a function of the evaporation temperature, 
%\begin{equation}
 %   \left(\frac{\Min}{1\rm g} \right) \simeq 9.93\times10^6 \left(1-\frac{3\gamma\lambda_c}{8} \right) \left(\frac{T_{\rm ev}}{1 \rm Gev} \right)^{-2/3}.
%\end{equation}
\begin{equation} \label{tevM}
    \left(\frac{\Min}{1\rm g} \right) \simeq 3.94\times10^8 \left(1-\frac{3\gamma\lambda_c}{8} \right) \left(\frac{T_{\rm ev}}{4 \rm Mev} \right)^{-2/3}.
\end{equation}
In the above expression, we consider $g_* = 106.75$ at the time of PBH formation. Equation (\ref{tevM}) clearly demonstrates the impact of accretion on the BBN limit for the minimum PBH mass. For example, if we set $\lambda_c =0$ (without  accretion), we recover the well known BBN bound on PBH initial mass \( M_{\rm in} \simeq 3.94 \times 10^8 \, \mathrm{g} \). On contrary, relativistic accretion (\( \lambda_c \simeq 10.4 \)) clearly shifts such bound to one order lower value, \( M_{\rm in} \simeq 8.6 \times 10^7 \, \mathrm{g} \).
%modifying their evaporation timeline and thereby tightly constraining the lower mass bound required to avoid conflict with BBN constraints. 
Our analytic estimates match well with the numerical ones illustrated in Fig. \ref{BBN_plot}, and further depicts the dependence of the initial PBH mass on $T_{\rm ev}$ for different accretion scenarios.
% {\red $\leftarrow$ This paragraph may be rewriten with more clarity.}

% To determine the shift in the lower mass limit, we solve the PBH mass evolution equations, Eqs.~(\ref{38}) and (\ref{39}), as a function of the background radiation temperature $T$. The scale factor $a$ scales inversely with $T$, i.e., $a \propto 1/T$, allowing us to express the scale factor ratio as $a/a_i = T_i/T$. Here, $T_i$ represents the initial temperature of the Universe when the PBH formed, which can be related to the initial PBH mass $M_i$ as:  
% \begin{equation}\label{initial_temp}
% T_i = M_i (4 \pi \gamma)^{\frac{1}{2}} \left( \frac{\pi^2}{90} g_{\ast}(T_i) \right)^{-\frac{1}{4}} \left( \frac{M_p}{M_i} \right)^{\frac{3}{2}}.
% \end{equation}  

% By solving the mass evolution equations in terms of $T_i/T$, we obtain the relationship between the initial PBH mass $M_i$ and the background radiation temperature at the time of PBH evaporation, $T_{\rm ev}$. This relationship is illustrated in Figure~\ref{BBN_constrain}. 

% where we have used the equation (\ref{17}) in order to take the accretion into account.

% The results demonstrate how accretion impacts the PBH mass evolution and consequently shifts the lower mass bound required to avoid disrupting BBN.

% \begin{figure}
%     \centering
%     \includegraphics[width=0.9\linewidth]{BBN_constrain.pdf}
%     \caption{Caption}
%     \label{BBN_constrain}
% \end{figure}

% \begin{figure}
%     \centering
%     \includegraphics[width=\columnwidth]{two_plots_BBN_mass_evolution.pdf}
%     \caption{\textbf{Left panel:} PBH mass ($M$) evolution as a function of scale factor ($a$) for initial masses of $10$ g (dashed) and $10^8$ g (solid), comparing relativistic (green) and non-relativistic (red) accretion along with Hawking evaporation.
% \textbf{Right panel:}  Initial PBH mass ($M_{\rm in}$) as a function of evaporation temperature ($T_{\rm ev}$), shown for three cases namely, blue, red, and green curves representing without, with non-relativistic, and with relativistic accretion respectively.}
%     \label{accretion}
% \end{figure}

\section{Improved constraints on DM}

\begin{figure}
    \centering
    \includegraphics[width=\columnwidth]{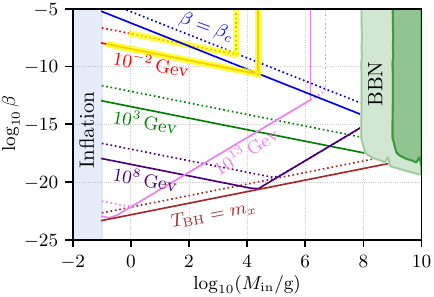}
    \caption{Variation of $\beta$ with $M_{\rm in}$ for various DM masses. Dotted lines represent scenarios with only Hawking evaporation, while the solid lines incorporate the effects of relativistic accretion as well.  
    % Dotted lines include only Hawking evaporation, while solid lines also account for relativistic accretion. 
    Shaded regions indicate constraints from inflation (light blue), BBN (green), improved BBN (light green), and Ly-$\alpha$ limits (yellow line).}
    \label{beta_vs_mi}
\end{figure}

PBH evaporation provides a compelling mechanism for explaining the observed DM abundance, $\Omega_{x ,0} h^2 \simeq 0.12$~\cite{Planck:2018vyg}. Assuming that the present DM is produced completely from the evaporating PBHs, and subsequently no additional entropy has been produced, the DM abundance can be expressed as \cite{Cheek:2021odj,Mambrini:2021cwd},
% PBH evaporation offers a compelling explanation for the observed DM abundance, $\Omega_{x ,0} h^2 \simeq 0.12$~\cite{Planck:2018vyg}. If we assume the present DM is produced completely from the evaporating PBHs, and subsequently no additional entropy has been produced, the DM abundance can be expressed as \cite{Cheek:2021odj,Mambrini:2021cwd},         %Therefore using equations (\ref{54}) and (\ref{initial_temp}), the expression for relic density becomes

\begin{align}
\Omega_{x,0}h^2 
%= & 3 (4 \pi \gamma)^{\frac{1}{2}} \left(\frac{\pi^2}{90} \right)^{\frac{3}{4}} \left( \frac{g_{\ast s}(T_{0})T_{0}^3}{\rho_{\text{crit},0}h^{-2}}  \right)\left(\frac{M_p}{M_{i}} \right)^{\frac{3}{2}}g_{\ast}(T_{i})^{-\frac{1}{4}} \notag \\
%& \times m_{DM} \beta N_{DM} \notag \\
    \simeq  1.7 \left(\frac{\gamma}{0.2} \right)^{\frac{1}{2}} \left(\frac{g_{\ast}}{106.75} \right)^{-\frac{1}{4}} \left(\frac{1\text{g}}{\Min} \right)^{\frac{3}{2}}
  \left(\frac{m_{x}}{1\text{Gev}} \right) \beta N_{x},
\end{align}
where $\beta = \rho_{\rm BH}^{\rm in} / \rho^{\rm in}$ is the PBH fraction defined as the ratio of the PBH energy density to the total background energy density at the time their formation. The total number of DM particles ($N_{x}$) produced from evaporation of PBH can then be computed using Eq.~(\ref{N_dm}) for a given DM mass $m_{x}$. For our present purpose we consider scalar dark matter particle.
% the initial PBH fraction $\beta = \rho_{\rm PBH}^i / \rho^i$, which \st{the} is the ratio of PBH energy density at the time of formation, and the background total energy density at that time. The total number of DM particles $N_{x}$ produce from PBH evaporation can now be computed using Eq.~(\ref{N_dm}), for a given DM mass $m_{x}$. 
For $m_{x}\lesssim T_{\rm BH}^{\rm acc}$, we have,
\begin{equation}
    \beta \leq 7 \times 10^{-10} \left( 1-\frac{3 \lambda_c \gamma}{8} \right)^2 \left( \frac{\Min}{1 \rm g} \right)^{-1/2} \left( \frac{1 \rm Gev}{m_{x}} \right)
\end{equation}
and for $m_{x} > T_{\rm BH}^{\rm acc}$,
\begin{equation}
    \beta \leq 7 \times 10^{-36}  \left( \frac{\Min}{1 \rm g} \right)^{3/2} \left( \frac{1 \rm Gev}{m_{x}} \right)^{-1} .
\end{equation}
Here, the equality represents the present day DM relic abundance shown in Fig.~\ref{beta_vs_mi} using solid brown, cyan, green and magenta lines. The appreciable departure from the standard cases without accretion is indeed observed particularly for dark matter masses $m_x < T_{\rm BH}^{\rm acc}$ as shown in respective dotted lines. 
%, ication in the $\beta - M_{i}$ relation due to general relativistic accretion, 
%as clearly illustrated by the solid line in Fig.~~\ref{beta_vs_mi}. 
Blue lines correspond to critical $\beta_c \equiv T_{\rm ev}/T_{\rm acc}$, above which we have matter dominated universe, and $\beta$ becomes independent of $M_{\rm in}$ represented by Vertical lines. For comparison, we also recover the relation in the absence of accretion, shown by the dotted lines.

An important constraint on the DM parameter space comes from the warm dark matter (WDM) bound. PBHs can emit DM particles with high momenta, which may remain relativistic until structure formation, conflicting with observations. The present day average DM velocity, expressed in terms of the average DM momentum $\langle p_{x} \rangle$, is given by
\begin{equation}
v_0 = \frac{a_{\rm ev}}{a_0} \frac{\langle p_{x} \rangle}{m_{x}} = \frac{a_{\rm ev}}{a_i} \frac{1}{1+z_{\rm eq}} \frac{T_{\rm eq}}{T_{\rm acc}} \frac{\langle p_{x} \rangle}{m_{x}},
\end{equation}
where $a_{\rm ev}$ and $a_0 = 1$ denote the scale factors at PBH evaporation and the present epoch, respectively. Here, $z_{\rm eq} \approx 3400$ is the redshift at matter-radiation equality, and the corresponding radiation temperature is $T_{\rm eq} \simeq 0.8 \times 10^{-9}$ GeV. For light DM particles $m_{x} \ll T_{\rm BH}^{\rm in}$, the average momentum is approximately $\langle p_{x} \rangle \sim T_{\rm BH}^{\rm in}$. The present-day velocity of WDM is given by \cite{Bode:2000gq}
%%%%%%%%%%%%%%%%%%%%%%%%%%%%%%%%%%%%%%%%%%%%%
\begin{equation}
v_{\rm WDM} \simeq 3.9 \times 10^{-8} \left( \frac{\rm KeV}{m_{\rm WDM}} \right)^{\frac{4}{3}}.
\end{equation}
%%%%%%%%%%%%%%%%%%%%%%%%%%%%%%%%%%%%%%%%%%%%%%%%%%
In our present analysis, we impose the lower bound on the WDM mass to be, $m_{\rm WDM} > 3.3 \, \rm KeV$ \cite{Viel:2013fqw}. This, in turn, constrains the DM mass as \cite{Haque:2023awl,Barman:2024iht}
\begin{equation}
\frac{m_{x}}{\rm GeV} \geq 8.1 \times 10^7 \left(\frac{m_{\rm WDM}}{\rm keV} \right)^{\frac{4}{3}} \left(\frac{M_{\rm acc}}{M_p}\right)^{\frac{1}{2}}.
\end{equation}
This constraint is shown in yellow in Fig.~\ref{beta_vs_mi}. Another important constraint arises from inflation. As discussed earlier, the Hubble parameter at the time of PBH formation is related to its initial mass by $H_{\rm in} = 4\pi \gamma M_p^2 M_{\rm in}^{-1}$. At the end of inflation, the Hubble parameter typically reaches a minimum value of $H_{\rm end} \sim 10^{14} \,\mathrm{GeV}$~\cite{ParticleDataGroup:2024cfk}, which imposes a lower bound on the PBH mass of approximately $M_{\rm in} \gtrsim 0.1 \,\mathrm{g}$. This constraint is represented by the light blue shaded region in Fig.~\ref{beta_vs_mi}.

To account for the effects of evaporation products on BBN within the mass range $10^8 \,\mathrm{g} \lesssim M_{\rm BH} \lesssim 10^{13} \,\mathrm{g}$, we follow the framework of Ref.~\cite{Keith:2020jww}, which models the photodisintegration rates of light nuclei. This constraint is represented by the green shaded region in Fig.~\ref{beta_vs_mi}. We further incorporate our modified evaporation history, which includes the impact of enhanced accretion, shown by the light green shaded region in the same figure.

%%%%%%%%%%%%%%%%%%%%%%%%%%%%%%%%%%%%%%%%%%%%%%%%%%%%%%%%%%%%

\section{Constrains on evaporating PBHs}

PBHs with masses below approximately $10^{15}\,\rm g$ are expected to have completely evaporated via Hawking radiation by the present epoch in the absence of accretion effects \cite{Carr:2020gox,Carr:2009jm}. However, incorporating accretion into the PBH evolution notably extends their lifetimes, thereby modifying the mass bound for surviving PBHs today.

\begin{figure}[t!]
    \centering
    \includegraphics[width=\linewidth]{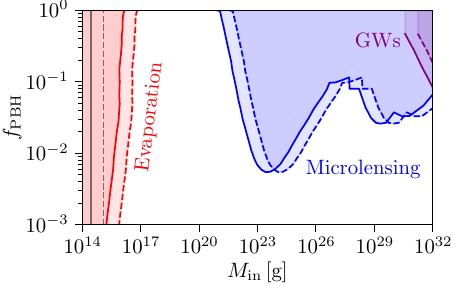}
    \caption{Constraints on the PBH dark matter fraction $f_{\rm PBH}$ vs. initial mass $M_{\rm in}$ are shown. The dashed curve includes only Hawking evaporation; the solid curve also accounts for relativistic accretion. Shaded regions denote bounds from evaporation (red), microlensing (blue), and GWs (purple). Black lines mark the critical mass with (solid) and without (dashed) accretion.}
    \label{f_PBH}
\end{figure}

Assuming that PBHs evaporate at the present cosmic time $t_0 \simeq 4.53 \times 10^{17}\, \rm s $, their lifetime is approximately given by $\tau_{\rm BH} = t_0 - t_{\rm acc} \simeq t_0$. Using this and the PBH lifetime relation in Eq.~(\ref{PBH_lifetime}), the modified initial mass for PBHs evaporating today is estimated as,
\begin{equation}
      M^0_{\rm in} \simeq \left( 3 \epsilon M_p^4 t_0 \right)^{\frac{1}{3}} \left(1 - \frac{3\gamma \lambda_c}{8} \right)  \simeq 1.2 \times 10^{15} \rm g  \left(1 - \frac{3\gamma \lambda_c}{8} \right) .
\end{equation}
This immediately modifies the initial PBH mass limit of approximately $\simeq 2.72 \times 10^{14}\, \rm g$, below which all PBHs would have fully evaporated by now. In the absence of relativistic accretion ($\lambda_c = 0$), this limit is about an order of magnitude higher, at $1.24 \times 10^{15}\,\rm g$. The inclusion of accretion thus introduces a new threshold, naturally shifting the parameter space of present-day PBH abundance, parametrized by $f_{\rm PBH}(M) = \Omega_{\rm PBH} / \Omega_{\rm DM}$, as a function of their initial mass $M_{\rm in}$.
% This immediately modifies the initial PBHs mass limit $\simeq 2.72 \times 10^{14}\,\rm g$ below which all PBH would have evaporated by now. Note that without the relativistic accretion $(\lambda_c = 0)$ such limit is one order higher $1.24 \times 10^{15}\,\rm g$. This accretion induced new threshold limit naturally shifts the parameter space of present day PBHs abundance parametrized by $f_{\rm PBH}(M) = {\Omega_{\rm PBH}}/{\Omega_{\rm DM}}$ versus their initial mass $M_i$. 
%Defining $f_{\rm PBH}(M)$ as the fraction of DM composed of PBHs of mass $M$, one can express:
%\begin{equation}
%    f_{\rm PBH}(M) = \frac{\Omega_{\rm PBH}}{\Omega_{\rm DM}},
%\end{equation}
Here, $\Omega_{\rm PBH}$ and $\Omega_{\rm DM}$ denote the present density parameters of PBHs and total DM, respectively. The constraints on $f_{\rm PBH}(M)$ arise from several observational channels~\cite{Oncins:2022ydg,Carr:2016drx,Green:2020jor,Carr:2020xqk}, including gravitational lensing \cite{Niikura:2019kqi,Niikura:2017zjd}, accretion-induced emissions \cite{Manshanden:2018tze,Ali-Haimoud:2016mbv}, extragalactic gamma-ray backgrounds~\cite{Carr:2016hva}, and gravitational wave observations~\cite{Nitz:2021vqh,Kavanagh:2018ggo}. Since accretion alters the mass evolution of PBHs, it induces a noticeable change in the viable parameter space, as illustrated in Fig.~\ref{f_PBH} using solid curves. This evidently leads to the modified mass range within $10^{16}\,{\rm g}$ to $ 10^{21}\,{\rm g}$ are still allowed to serve as the sole DM candidate.

%%%%%%%%%%%%%%%%%%%%%%%%%%%%%%%%%%%%%%%%%%%%%%%%%%%%%%%%%%%%%%%%

% $M_{original}\approx 1.23\times 10^{15}\,\rm g$, $M_{New} \approx 2.72\times 10^{14}\,\rm g$
\section{Gravitational Waves from PBH Evaporation}

PBHs evaporate via Hawking radiation, emitting gravitons that contribute to a stochastic gravitational wave (GW) background. This GW signal offers a unique window into PBH physics and the early universe. Accurate modeling requires going beyond idealized blackbody assumptions using tools like \texttt{BlackHawk} \cite{Arbey:2019mbc}, which accounts for greybody factors, spin effects, and realistic emission spectra.

The instantaneous GW energy density is given by~ \cite{Anantua:2008am,Ireland:2023avg},
\bea
\frac{d\rho_{\text{GW}}}{dt\,d\omega} = n_{\text{BH}}(t)\, \frac{\omega}{2\pi}\, Q_{\text{GW}}(t, \omega,M_{\rm acc}),
\eea
where $Q_{\text{GW}}$ is the graviton flux computed numerically. The GW energy density at evaporation is obtained by integrating over the PBH lifetime; redshifting this result gives the present day spectrum as,
%\bea
%\frac{d\rho^{\rm ev}_{\text{GW}}}{d\ln\omega_{\rm ev}} = n_{\text{BH}}^{\rm i} \left(\frac{a_{\rm i}}{a_{\rm ev}}\right)^3 \frac{\omega_{\rm ev}^2}{2\pi} \int_{t_i}^{t_{\rm ev}} dt\, \frac{a_{\rm ev}}{a(t)}\, Q_{\text{GW}}\left(t, \frac{\omega_{\rm ev} a_{\rm ev}}{a(t)}\right).
%\eea
\bea
\Omega_{\rm GW}(f) = \frac{\rho_{\rm BH}^{\rm in}}{M_{\rm in}\,\rho_{\rm c}^0} \frac{\omega_0^2}{2\pi} \frac{a_{\rm in}^3}{a_0^2} \int_{t_{\rm acc}}^{t_{\rm ev}} \frac{dt}{a(t)}\, Q_{\text{GW}}\left(t, \frac{\omega_0 a_0}{a(t)}\right),
\eea
where $ \rho_{\rm c}^0 $ and $ a_{\rm 0} $ denote the critical energy density and scale factor measured today, respectively.
\begin{figure}[t]
    \centering
    \includegraphics[width=0.9\linewidth]{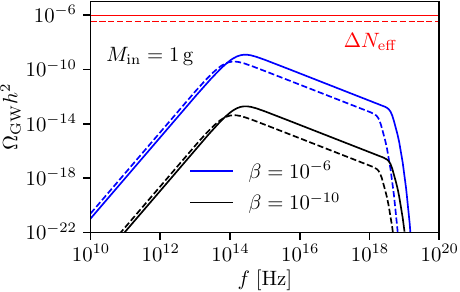}
    \caption{ $\Omega_{\rm GW}$ is shown versus frequency. The solid line includes both Hawking evaporation and relativistic accretion; the dashed line shows only Hawking emission. Red curves indicate $\Delta N_{\rm eff} = 0.17$ (Planck + BAO)~~\cite{Planck:2018vyg} and $\Delta N_{\rm eff} = 0.06$ (CMB-S4)~~\cite{Abazajian:2019eic} constraints.}
    \label{GW_PBH}
\end{figure}
%%%%%%%%%%%%%%%%
Relativistic accretion extends PBH lifetimes by modifying the mass loss rate, which in turn shifts the peak frequency and amplitude of the resulting GW spectrum (see Fig.~~\ref{GW_PBH}). While our primary results and the analytical expressions derived here rely on the assumption of a monochromatic PBH mass function, we note that an extended mass function, $\psi(M_{\rm in})$, would modify the resulting spectrum. In this generalized case, the total observable GW spectrum is the superposition of all the monochromatic spectra, weighted by the mass distribution $\psi(M_{\rm in})$. Since the peak frequency scales directly with mass~\cite{Ireland:2023avg}, the primary effect would be a broadening of the spectrum. Consequently, the single, sharp peak corresponding to a single $M_{\rm in}$ would be ``smeared out'' over a range of frequencies defined by the width and shape of the PBH mass function itself.

Detecting these high-frequency GWs is challenging but potentially feasible with next-generation detectors, including laser interferometers (1–10 kHz)~~\cite{Ackley:2020atn}, levitated sensors (10–100 kHz)~~\cite{Aggarwal:2020olq}, magnetic conversion (10 GHz)~~\cite{Li:2009zzy,Ito:2019wcb}, and inverse Gertsenshtein effect devices ($10^{14}–10^{18}$ Hz)~~\cite{Ejlli:2019bqj}. Axion experiments such as JURA~\cite{Beacham:2019nyx}, OSQAR II \cite{OSQAR:2015qdv}, and CAST ~\cite{CAST:2017uph} may be repurposed for graviton detection, providing novel probes of PBH accretion and early-universe dynamics.

\section{Conclusions}

Primordial Black Holes (PBHs) have become an integral part of theoretical model building for the early Universe and dark matter. In this paper, we revisit the cosmological evolution of PBHs by incorporating two key physical processes: Hawking evaporation and accretion. In contrast to previous studies that modeled accretion using a pseudo-Newtonian approximation, which typically yields negligible mass growth, we formulate the accretion dynamics within a fully general relativistic framework. We demonstrate that relativistic corrections, crucial in the dense plasma of the early universe, significantly enhance the accretion rate, leading to a phase of rapid, runaway PBH mass growth on timescales much shorter than those of Hawking evaporation. This dynamic interplay imposes a new and stronger lower bound on the PBH mass, $M_{\rm in} \leq  6.9\times 10^{7}\, \rm g$, as PBHs forming below this mass would grow so efficiently that their subsequent, more powerful evaporation injects excessive energy into the primordial plasma, disrupting the successful predictions of Big Bang Nucleosynthesis (BBN).

This accretion-dominated evolution has several profound cosmological implications that we have explored. We analyze the consequences of this enhanced accretion on Dark Matter (DM) production via Hawking radiation and find a noticeable shift in the allowed parameter space spanned by the PBH abundance and DM mass, modifying constraints relative to previous results~\cite{Cheek:2021odj,Gondolo:2020uqv}. This modification arises because the rapid mass increase alters the evaporation lifetime and temperature of the PBHs ($T_{\rm BH} \propto 1/M_{\rm BH}$), thereby changing the production efficiency and viable parameter space for DM particles generated via Hawking radiation. As a PBH grows, it ceases to produce massive DM particles sooner than a non-accreting PBH would, thus altering the final relic abundance.

Furthermore, we determine that the minimum PBH mass capable of surviving to the present day is reduced to $\sim 2 \times 10^{14}\, \rm g$, an order of magnitude below existing estimates \cite{Carr:2009jm,Carr:2020xqk}. PBHs with masses above this threshold could themselves constitute viable DM candidates. This result significantly revises the minimum mass for which a PBH can be considered a potential dark matter constituent today, effectively widening the available parameter space. Notably, our work reinforces the viability of the broad mass window between $10^{16}\, \rm g$ and $10^{21}\, \rm g$, which remains observationally unconstrained and allows PBHs to account for the entirety of DM. Finally, we compute modifications to the high-frequency gravitational wave (GW) spectrum arising from this accretion-driven PBH mass growth. 

In summary, our work underscores the critical importance of incorporating a fully general relativistic treatment for PBH accretion. By doing so, we have refined key cosmological constraints related to BBN and dark matter, and we have highlighted a clear, potentially observable signature in the GW background. Future multi-messenger explorations of the early Universe will be crucial for testing the accretion-driven scenario presented herein. Important directions in which our present work can be extended could to take into account Kerr-black holes, impact of thermal feedback on the accretion process etc.

\section{ACKNOWLEDGEMENTS}
MRH acknowledges ISI Kolkata for providing financial support through the Research Associateship. JK would like to thank Gargi Sen and Samik Mitra for their help and useful discussions. DM gratefully acknowledges support from the Science and Engineering Research Board (SERB), Department of Science and Technology (DST), Government of India (GoI), through the Core Research Grant CRG/2020/003664.

\end{document}